\documentclass[12pt]{article}
\usepackage{latexsym}

\newcommand{\xxx}[1]{ [#1]}
\def\thefootnote{\fnsymbol{footnote}}
\newcommand{\vev}{{\it vev}}

\def\r{\right|}
\def\l{\left.}
\def\[{\left [}
\def\]{\right ]}
\def\({\left (}
\def\){\right )}

\def\ux{$U(1)_X$}
\def\ua{$U(1)_a$}

\def\im{{\rm Im}}

\newcommand{\beq}{\begin{equation}}
\newcommand{\eeq}{\end{equation}}
\newcommand{\bea}{\begin{eqnarray}}
\newcommand{\eea}{\end{eqnarray}}
\def\bF{\bar{F}}

\def\del{\delta}

\newcommand{\myref}[1]{(\ref{#1})}
\newcommand{\beqa}{\begin{eqnarray}}
\newcommand{\eeqa}{\end{eqnarray}}
\newcommand{\nnn}{ \nonumber \\ }

\newcommand{\Zbf}{{{\bf Z}}}

\newcommand{\uone}{$U(1)$}

\newcommand{\myvev}[1]{{\langle #1 \rangle}}
\newcommand{\bigvev}[1]{{\left\langle #1 \right\rangle}}

\newcommand{\half}{{1 \over 2}}

\def\eee{\nonumber \\ &=&}

\begin{document}

\begin{titlepage}

\hfill   LBNL-56687

\hfill   UCB-PTH-04/35

\hfill   hep-th/0412079

\hfill \today

\begin{center}

\vspace{18pt}
{\bf R-parity from the heterotic string}\footnote{This
work was supported in part by the
Director, Office of Science, Office of High Energy and Nuclear
Physics, Division of High Energy Physics of the U.S. Department of
Energy under Contract DE-AC03-76SF00098, and in part by the National
Science Foundation under grant PHY-0098840.}

\vspace{18pt}

Mary K. Gaillard\footnote{E-Mail: {\tt MKGaillard@lbl.gov}}
\vskip .01in
{\em Department of Physics, University of California 
and \\ Theoretical Physics Group, Bldg. 50A5104,
Lawrence Berkeley National Laboratory \\ Berkeley,
CA 94720 USA}\vskip .03in

\vspace{18pt}

\end{center}

\begin{abstract}
In T-duality invariant effective supergravity with gaugino condensation
as the mechanism for supersymmetry breaking, there is a residual discrete
symmetry that could play the role of R-parity in supersymmetric
extensions of the Standard Model.
\end{abstract}

\end{titlepage}

\newpage
\renewcommand{\thepage}{\roman{page}}
\setcounter{page}{2}
\mbox{ }

\vskip 1in

\begin{center}
{\bf Disclaimer}
\end{center}

\vskip .2in

\begin{scriptsize}
\begin{quotation}
This document was prepared as an account of work sponsored by the United
States Government. Neither the United States Government nor any agency
thereof, nor The Regents of the University of California, nor any of their
employees, makes any warranty, express or implied, or assumes any legal
liability or responsibility for the accuracy, completeness, or usefulness
of any information, apparatus, product, or process disclosed, or represents
that its use would not infringe privately owned rights. Reference herein
to any specific commercial products process, or service by its trade name,
trademark, manufacturer, or otherwise, does not necessarily constitute or
imply its endorsement, recommendation, or favoring by the United States
Government or any agency thereof, or The Regents of the University of
California. The views and opinions of authors expressed herein do not
necessarily state or reflect those of the United States Government or any
agency thereof of The Regents of the University of California and shall
not be used for advertising or product endorsement purposes.
\end{quotation}
\end{scriptsize}

\vskip 2in

\begin{center}
\begin{small}
{\it Lawrence Berkeley Laboratory is an equal opportunity employer.}
\end{small}
\end{center}

\newpage
\renewcommand{\thepage}{\arabic{page}}
\setcounter{page}{1}
\def\thefootnote{\arabic{footnote}}
\setcounter{footnote}{0}

In the context of the weakly interacting heterotic string, T-duality
invariant effective supergravity Lagrangians have been
constructed~\cite{bgw,ggm} for supersymmetry breaking by condensation
in a hidden sector.  The T-duality of these models
assures\footnote{This result is assured only if the constraint on the
gaugino condensate that follows from the Yang-Mills Bianchi identity
is imposed.} that the T-moduli are stabilized at self-dual points with
vanishing vacuum values (\vev's) for their auxiliary fields.  Thus
supersymmetry breaking is dilaton dominated, thereby
avoiding\footnote{The effects of quadratically divergent loop
corrections~\cite{clm} will be examined elsewhere~\cite{gnq}.} a
potentially dangerous source of flavor changing neutral currents
(FCNC).  Another consequence of this result is that there is a
residual discrete symmetry that might play the role of R-parity in the
minimal supersymmetric extension (MSSM) of the Standard Model (SM).

The heterotic string is perturbatively invariant~\cite{mod} under
transformations on the T-moduli, that, in the class of models
considered here, take the form\footnote{We neglect mixing~\cite{mix}
among twisted sector fields of the same modular weights $q^A_I$ with
mixing parameters that depend on the integers $a^I,b^I,c^I,d^I$.}
\bea T^I &\to& {a^I T^I - i b^I \over ic^I T^I + d^I}, \nnn \Phi^A
&\to& e^{-\sum_I q_I^A F^I} \Phi^A,\nnn a^I d^I - b^I c^I &=& 1,
\qquad a^I,b^I,c^I,d^I \in \Zbf \qquad \forall \quad I=1,2,3, \nnn F^I
&=& \ln \( i c^I T^I + d^I \),
\label{mdtr}
\eea
and under which the K\"ahler potential and superpotential 
transform as 
\beq K\to K + F + \bF, \qquad W\to e^{- F}W, \qquad F = \sum_I
F^I.\eeq
The self-dual vacua $T_{s d}$, namely
\beq  \myvev{t^I} = 1 \qquad {\rm or}\qquad e^{i\pi/6}, \eeq
are invariant under \myref{mdtr} with
\beq b^I = - c^I = \pm 1, \qquad a^I= d^I = 0\quad {\rm or}\quad
\cases{a^I = b^I,\; d^I = 0\cr d^I = c^I,\;a^I = 0\cr} , \qquad
F^I =  n i{\pi\over2}\quad{\rm or}\quad n i{\pi\over3}\label{z2z3}.\eeq
So for three moduli we have a symmetry under $G_R = Z_2^m\otimes
Z_3^{m'},\; m + m' = 3$.  The gaugino condensates $u$ [and matter
condensates $\myvev{e^KW(\Pi)}\propto u$; see \cite{bgw}] that get
\vev's break this further to a subgroup with
\beq i\im F = F = 2n i\pi,\label{subg}\eeq
under which $\lambda_L\to e^{-{i\over2}\im F}\lambda_L = \pm
\lambda_L$; we would identify the case with a minus sign with
R-parity.  This subgroup also leaves invariant the soft
supersymmetry-breaking terms in the observable sector, if no other
field gets a \vev\, that breaks it.  For example if the $\mu$-term
comes from a superpotential term $H_u H_d \Phi$, with the \vev\,
$\myvev{\phi = \l\Phi\r}\ne 0$ generated at the TeV scale, the
symmetry could be broken further to a subgroup $R\in G_R$ such that
$R\Phi = \Phi$.  On the other hand if the $\mu$-term comes from a
K\"ahler potential term generated by invariant \vev's above the scale
where the moduli are fixed there would be no further breaking until
the Higgs get \vev's; with the residual R-parity satisfying $R H_{u,d}
= H_{u,d}$.

Given the transformation property
\beq \eta(i T^I)\to e^{i\del_I}e^{\half F(T^I)}\eta(i T^I),\qquad F(T^I) =
F^I,\qquad \del_I = \del_I(a^I,b^I,c^I,d^I),\label{phase}\eeq
of the Dedekin $\eta$-function, superpotential terms of the form
\beq W = \prod_A\Phi^A\prod_I\eta(i T^I)^{2\(\sum_A q^A_I -
1\)},\label{wterm}\eeq
would be covariant under \myref{mdtr} if the moduli independent
phases~\cite{lust} satisfied $\del_I = 2n^I i\pi$.  However it is
easy to see that this is not the case for the transformations that
leave fixed the self-dual points $T_{sd}$:
\beq \eta(i T_{sd})\to \eta(i T_{sd})\ne e^{\half
F(T_{sd})}\eta(i T_{sd})\label{sdt}.\eeq
It follows from T-duality that this phase can be reabsorbed~\cite{fer}
into the transformation properties of the twisted sector fields.
Consider for example a $Z_3$ orbifold with twisted sector fields $T^A$
and $Y^{A I}$ with modular weights
\beq q_I^A = \({2\over3},{2\over3},{2\over3}\),\qquad
\(q_I^{A J}\)_Y = \({2\over3},{2\over3},{2\over3}\) + \del_I^J.\eeq
The untwisted sector fields $U^{A I}$ have modular weights
\beq \(q_I^{A J}\)_U = \del_I^J.\eeq
Then since allowed superpotential couplings are of the
form~\cite{hv} $T^{3p}U^q$, invariance under \myref{sdt}, subject to
\myref{subg}, requires that $\prod\Phi^A$ in \myref{wterm} gets, in
addition to the phases implicit in \myref{mdtr}, an overall phase
factor $\del_{sd}$ that satisfies
\beq i\del_{sd} = \sum_I(2p + n^I)F(T^I_{sd}) = \sum_I n^I
F(T^I_{sd}) + 4ik\pi,\qquad k\in \Zbf\label{kwt}\eeq
where $n^I$ is the number of $U^{A I}$ and $T^{A I}$ factors in
$\prod\Phi^A$.  More generally, T-duality implies that the allowed
terms in the superpotential must be such that there is a choice of
phases $\del^A$ that makes it covariant if the transformation of
$\Phi^A$ in \myref{mdtr} is modified to read
\beq \Phi^A \to e^{i\del^A - \sum_I q_I^A F^I}
\Phi^A.\label{deltr}\eeq
For example in~\cite{fer} trilinear terms $W\sim U_1U_2U_3,(T)^3,$
where $T= T^A$ were considered; all such terms would be covariant
provided $\del^U = 0$, $\del^T = - {2\over3}\del = -
{2\over3}\sum_I\del_I$, .  If we further impose $\del^{Y^J} = -
{2\over3}\del - 4\del_I$, then the monomials
\beq U_1U_2U_3,\qquad T^3\Pi^p,\qquad U_I Y^I T^2\Pi^p,\qquad
U_I Y^I U_J Y^J T\Pi^p, \qquad U_I Y^I U_J Y^J U_K Y^K\Pi^p,\label{mons}\eeq
where $\Pi = Y^1Y^2Y^3,$ can be used to construct covariant
superpotential terms $W_\alpha$, by multiplying them by powers of
$\eta_I = \eta(i T^I)$ such that the overall monomials
$W_\alpha$ are modular covariant with modular weights $q^\alpha_I =
1$. Further operators can be constructed by multiplying these by
invariant operators; for example $\Pi\eta^6,\; \eta = \prod_I\eta_I$
is invariant. In addition, invariant operators of the form
\beq \eta^{2m}\prod_{i=1}^m W_{\alpha_i}, \qquad 2m\del = 2\pi n,
\label{invmons} \eeq
can be constructed since $\delta/\pi$ is a rational
number.\footnote{The group \myref{mdtr} of duality transformations on
$T^I$ is generated~\cite{SW} by $T^I\to1/T^I$ with $\del(1,1,0,1) =
\pi/4$, and $T^I\to T^I -i$ with $\del(0,1,-1,0)=\pi/12$.}  These
couplings are consistent with the selection rules~\cite{hv}.  They are
further restricted by additional selection rules and gauge
invariance. The invariant operators \myref{invmons} can also be used
to construct terms in the K\"ahler potential.  For the subgroup
defined by \myref{z2z3} and \myref{subg}, $i\del = - \half F = -
in\pi$, the superpotential is invariant, as are the monomials in
\myref{mons}, so any product of them in could appear in the effective
superpotential or K\"ahler potential, {\it e.g.}, through quantum
corrections and/or integrating out massive fields, in the effective
theory below the scale where the T-moduli are fixed and supersymmetry
is broken, with possibly additional \vev's that are invariant under
$G_R$ generated at that scale.

Superpotential terms of dimension three will be generated from higher
order terms when some fields acquire \vev's.  In models with an
anomalous \ux, there is a Green-Schwarz counterterm in the form of a
D-term~\cite{UXR} that leads to the breaking of a number $m$ of
\uone\, gauge factors when $n\ge m$ fields $\Phi^A$ acquire \vev's.
T-duality remains unbroken~\cite{GG02b}, but the modular weights are
modified by going to unitary gauge in a way that keeps modular
invariance manifest. For example in minimal models with $n=m$:
\beq \Phi^M \to \Phi'^M,\qquad q^M_I \to q'^M_I = q^M_I - \sum_{A
a}q^M_a Q^a_A q^A_I, \qquad \sum_A Q^a_A q^A_b = \del^a_b, \qquad
\sum_a Q^a_A q^B_a = \del^B_A.\label{newq}\eeq
This has the effect of making the eaten chiral supermultiplets
modular invariant in the superhiggs mechanism.  Then for a term 
in the superpotential \myref{wterm} with some $\bigvev{\Phi^A}\ne0$:
\bea W &=& \prod_M\Phi^M\prod_B\bigvev{\Phi^A}
\prod_I\eta(i T^I)^{2\(\sum_M q^M_I + \sum_B q^A_I - 1\)}\eee
\prod_M\Phi'^M\prod_B\bigvev{\Phi^A}\prod_I\eta(i T^I)^{2\(\sum_M q'^M_I
- 1\)}, \label{wterm2}\eea
because $W$ is also \ua\, invariant: $\sum_M q^M_a + \sum_A q^A_a =
0$.  In order to make T-duality fully manifest below the
\uone-breaking scale, we have to redefine the transformation
\myref{deltr} by including a global \ua\, transformation such that
$\Phi^A$ is fully invariant, and
\beq \Phi'^M \to e^{i\del'^M - \sum_I q'^M_I F^I} \Phi^M, \qquad
\del'^M = \del^M - \sum_{A a}q^M_a
Q^a_A\del^A.\label{deltr2}\eeq

{\it A priori} we expect that $\myvev{\Phi^A}\sim .1$, so that
couplings arising from high dimension operators in the superpotential
are suppressed.\footnote{The factors multiplying these terms can in
fact be rather large~\cite{joel}.}  We would like to have one large coupling
($Q_3,T^c,H^u$) which should correspond to one of the dimension three
operators in \myref{mons}.  Most models studied~\cite{mods,fiqs} have
quark doublets in the untwisted sector. In this case we should take
$T^c$ and $H^u$ in the untwisted sector as well, and
require\footnote{This requirement in satisfied in the FIQS
model~\cite{fiqs}.} $q^{Q_3}_a + q^{T^c}_a + q^{H^u} = 0$.  That is,
if we identify the $Q_I$ generation index with the moduli index, we
can have, {\it e.g.}, $T^c = T^c_2$, $H^u = H^u_1$.  Then to suppress
the $Q_2 C^c H^u$ and $Q_2 U^c H^u$ couplings we require
$C^c,U^c\notin U_3$, so one of these must be in the untwisted sector
$T$. Since these generally have different \uone\, charges from the
untwisted sector fields, to avoid a possible D-term induced
flavor-dependence of the squark masses in the first two generations,
we also take both $U^c$ and $C^c$ in $T$.

As an example (that turns out not to produce the desired R-parity)
consider the FIQS model~\cite{fiqs}, with the $\phi^A$ vacuum studied
in~\cite{ggm}. Then $D^c,S^c,B^c,H^u\in T$.  To generate all the
known Yukawa's ($Q T^2,Q H^u T$) it follows from \myref{mons} that at
least three $Y^I$ with different indices $I$ have to have \vev's.  If
all the \vev's are generated by D-term breaking, we have to choose the
three-fold version~\cite{ggm} of the ``minimal'' FIQS model with
$\myvev{Y_1^{1,2,3}}\ne 0$.  Defining
\beq \zeta^M_A = \sum Q^a_A q^M_a, \qquad \zeta^M = \sum_A\zeta^M_A,\eeq
we have $\zeta^M_{Y_1} = {2\over3}\zeta^M - \sqrt{3\over2}q^M_X$.  Then
after the redefinitions \myref{newq} and \myref{deltr2} with $i\del_I = -
\half F^I$, we have
\beq T'^M\to e^{\[{1\over3}(\zeta^M - 1) +
\sqrt{3\over2}q^M_X\]F}T'^M.\label{fiqs}\eeq
For all the possible MSSM candidates the $T^M$ have  
$\sqrt{3\over2}q^M_X = {1\over3}n^M$ and we just get
\beq T'^M\to e^{{1\over3}(\zeta^M - 1 + n^M)F}T'^M.\eeq
Then, using \myref{subg}, trilinear terms $T^3$ with fixed $n^M$
satisfy
\beq \prod_{i = 1}^3T'^{M_i} \to e^{{1\over3}\sum_{j=1}^3\zeta^{M_j}F}
\prod_i T'^{M_i}.\label{Ttr}\eeq
Taking the $T^{M_i}$ in \myref{Ttr} to be the FIQS supermultiplets
$U^c=u_2\in T$ and any two of $D^c,S^c,B^c = d_{1,2}\in T,$ we have
$n^{M_i} = -1,\;\sum_j \zeta^{M_j} = 0$, so we cannot forbid baryon
number violating couplings.  We can nevertheless ask if we get any
interesting restrictions. It turns out that for SM gauge invariant
trilinear couplings we get $\sum_j\zeta^{M_j\ne\ell_5} = n$,
$\sum_j\zeta^{M_j\ne\ell_5} + \zeta^{\ell_5}= {n\over2}$, and after
imposing \myref{subg}, aside from couplings involving $\ell_5$,
everything drops out except the original T-duality transformation on
the untwisted fields; if for an operator $O$, $O\to\eta(O)O$, we have
\bea \eta(G_i G_j \ell_{k\ne5}) &=& 1,\qquad \eta(G_i G_j \ell_5) =
e^{{n\over3}i\pi}, \qquad \eta(Q_I d_i G_j) = e^{-F^I}, \nnn \eta(Q_I
u^1_J\tilde G^1_K) &=& e^{- F^I - F^J - F^K}, \qquad \eta(Q_I
u_2\tilde G_{k\ne1}) = e^{- F^I},\nnn \eta(Q_I u^1_J\tilde G_{k\ne1})
&=& \eta(Q_I u_2\tilde G^1_J) = e^{- F^I - F^J},\label{coup}\eea
Thus $L^2E^c$ is allowed unless $E^c = \ell_5$, in which case $L H_d
E^c$ is also forbidden, unless the symmetry is broken to $n = 6p$, in
which case both are allowed.  In order to have at least one $Q_I H_d
D^c$-type coupling for each $Q_I$ we need $F^I = 2in\pi\;\forall\; I$.
Then all couplings involving the $Q_I$ are allowed, including $Q_I L
D^c$, {\it etc.}  We can also look at candidate $\mu$-term couplings
$H_u H_d = \tilde G_i G_j$; these have $\zeta^{\tilde G_i} +
\zeta^{G_j} = m$, and
\beq \eta(G_i\tilde G^1_I) = e^{{m\over3}i\pi + F^I},\qquad
\eta(G_i\tilde G_{j\ne1}) = e^{{2m\over3}i\pi},\label{higgs}\eeq
so \myref{subg} has to be broken to a smaller subgroup when the
$\mu$-term is generated. 

Apart from the fact that the FIQS model doesn't give the correct
constraints, it is still interesting to see if one can get any
unbroken symmetry after the Higgs particles acquire \vev's.  In this
model the individual $\eta$'s are of the form $e^{2ni\pi{m\over33}}$,
except for $\ell_5$ where ${m\over33}\to {2m+1\over66}$, so the individual
\vev's of $H_{u,d}$ break the symmetry down to a subgroup with $n =
33p$ in \myref{subg}. If this is the only symmetry left the only
constraint on the couplings in \myref{coup} is to forbid $G_i
G_j\ell_5$. However, we can once again redefine the transformations
such that one Higgs is invariant and the other has the phase factor in
\myref{higgs}, and therefore both Higgs are invariant under the
subgroup left unbroken by the $\mu$-term.  Here we use the fact that
the couplings are invariant under electroweak hypercharge $Y$, and
redefine the transformation properties by
\beq \eta_M\to \eta_M\eta_{H_u}^{-2Y^M}\label{redef}.\eeq
Then $H_u$ with $Y^{H_u} =\half$ is invariant and the couplings that
were allowed/forbidden under the group left unbroken by the $\mu$-term
\myref{higgs} remain so.

Returning to the viability of the FIQS model, not all the \vev's have
to be generated at the \uone\, breaking scale.  For example after
condensation soft masses and A-terms are generated, but the Lagrangian
is $G_R$ invariant.  If just one more field gets a \vev, one can do
another redefinition as in \myref{deltr2} such that this field is
$G_R$ invariant, and the net effect must be the same.  If several
fields get \vev's and there is a residual subgroup $R$ that survives,
it must be possible to redefine all of them to be invariant, as above,
by exploiting surviving gauge symmetries at that scale, so in this
model it appears that generating the observed couplings does not admit
an R-symmetry that could forbid the unwanted ones.

Now we turn to a more general analysis, assuming the same assignments as
before for the MSSM fields, but with different \uone\, charges.  Then
the analogue of \myref{fiqs} is
\bea U'^M_J&\to& e^{{1\over3}\zeta^M F - \sum_I\zeta_I^{M J}F^I - F^J}
U'^M_J = \eta_{M J}U'^M_J,\nnn
  T'^M &\to& e^{{1\over3}(\zeta^M - 1)F -
 \sum_I\zeta_I^M F^I}T'^M = \eta_M T'^M,\label{gentr}\eea
where
\beq \zeta^M_I = \sum_A\zeta^M_{Y^A_I}, \qquad \zeta_A^Q + \zeta_A^{T^c}
+ \zeta_A^{H^u} = 0, \qquad \eta_{Q_3}\eta_{T^c}\eta_{H^u} = 1.\eeq
We also require $\eta_{H_u}\eta_{H_d} = 1$.  If $\tilde D^c =
D^c,S^c,B^c$ all have the same \uone\, charges, then
$\eta_{D^c}=\eta_{S^c}=\eta_{B^c}$.  Then in order to have at least
one coupling $Q_I\tilde D^c H_d$ for each value of $I$, we require
$\eta_{Q_I} = \eta_Q$ independent of $I$, and requiring at least one
coupling $Q_I C^c H_u$ implies $\eta_{U^c} = \eta_{C^c} = \eta_{T^c}$
if $\tilde U^c = U^c,C^c$ have the same \uone\, charges.  Then all
couplings of these two types are allowed.  Similarly if we assume the
lepton doublets $L$ and singlets $E^c$ have (two sets of) degenerate
\uone\, charges $q_a^L,q_a^{E^c}$, they also have degenerate
R-parities: $\eta_L,\eta_{E^c}$.  To forbid $L^2 E^c$, $L Q\tilde D^c$
and $L H_u$ we require $\eta_L\ne\eta_{H_d}$, and to forbid $\tilde
D^2\tilde U^c$, we require $\eta_{U^c}\ne (\eta_{D^c})^{-2}$ or
\beq \eta^2_{\tilde D^c}\eta_{\tilde U^c} =
\eta^{-3}_Q\eta_{H_u}\ne1.\eeq
If, as in the FIQS model,
the $Q_I$ all have the same \uone\, charges, the constraint that
they have the same R-charge implies that $F^I - F^J = 2n i\pi$.
Then since we also require $\sum_I F^I = 2m i\pi$, it is easy to
check that $F^I = 2n^I i\pi$, giving
\beq \eta_{M J} = e^{2i\pi\sum_I n^I\({1\over3}\zeta^M - \zeta_I^M\)},
 \qquad \eta_M = e^{2i\pi\sum_I n^I\[{1\over3}(\zeta^M - 1) -
 \zeta_I^M\]}.\label{gentr2}\eeq
We can find an R-parity provided there is some compactification for
which we can identify the particles of the SM in such a way that the
above constraints are satisfied.  With the above choices
[$Q_I,T^c,H_u\in U;\;\tilde U^c,\tilde D^c,L,E^c,H_d\in T$ and
degenerate \uone\, charges for fixed flavor in each sector], they take
the form
\bea \eta_Q &=& e^{2i\pi\beta}\eta_{H_d}^{-2},\qquad \eta_{\tilde U^c}
= \eta_{T^c} = e^{-2i\pi\beta} \eta_{H_d}^3, \qquad \eta_{\tilde D^c}
= e^{-2i\pi\beta}\eta_{H_d},\qquad\beta\ne {n\over3} \nnn \eta_{H_u}
&=& \eta_{H_d}^{-1}, \qquad \eta_L = e^{2i\pi\alpha}\eta_{H_d},\qquad
\eta_{E^c} = e^{-2i\pi\alpha}\eta_{H_d}^{-2}, \qquad 0< \alpha,\beta<
1.\label{phases}\eea
When the electroweak symmetry is broken, we redefine R-parity as
in \myref{redef} so that $\eta_{H_d} = 1$.

Other scenarios can be considered.  For example a $Q_3T^c H_u$
coupling is allowed if these are all in the twisted sector $T$, their
\uone\, charges sum to zero and there is no $\Pi$ factor in
\myref{mons} required by a further string symmetry.  In this case it
would be possible to have all quarks of the same flavor having the
same \uone\, charge.  It is in fact not necessary to have identical
\uone\, charges to assure equal masses for squarks and sleptons of the
same flavor, which is what is actually needed to avoid unwanted FCNC;
scalars $\phi^M$ with the same value of $\zeta^M$ have the same
masses, but could have different values\footnote{In the FIQS model
considered above, the MSSM candidates with the same flavor that are
degenerate in $\zeta^M$ are also degenerate in $\zeta^M_I$.} of
$\zeta^M_I$.  The viability of these scenarios in the context of
gaugino condensation favors vanishing or very small values of
$\zeta^M$ for SM particles, so the values of the $\zeta^M_I$ could be
the governing factors in determining R-parity.

To what extent have we achieved the conventional definition of
R-parity?  With appropriate choices of phases in \myref{phases} we
achieve the elimination of baryon and lepton number violating
couplings\footnote{Fast proton decay is avoided by eliminating either
one of these, but the $\tilde U^c(\tilde D^c)^2$ coupling by itself
would induce neutron-anti-neutron oscillations.} of dimension two or
three in the superpotential.  Higher dimension operators can generate
$B$ and $L$ violation, as is the case with the conventional definition
of R-parity. In the latter case the R-allowed dimension-four operator
$\tilde U^c\tilde U^c\tilde D^c E^c$ in the superpotential leads to
dimension-five operators in the effective Lagrangian that may be
problematic~\cite{hit} even if these couplings are Planck- or
string-scale suppressed, given the current bounds on the proton decay
lifetime. This problem easily evaded in the current context provided
\beq 3\beta + \alpha\ne n.\eeq
The stability of the lightest neutralino is assured at the same level
as proton stability since its decay products would have to include an
odd number of SM fermions and hence violate $B$ and/or $L$.

A more comprehensive examination of $Z_3$ orbifold models in this context
will be presented elsewhere~\cite{gg}.

\section*{Acknowledgments}
I wish to thank Joel Giedt for helpful input.  This work was supported
in part by the Director, Office of Science, Office of High Energy and
Nuclear Physics, Division of High Energy Physics of the
U.S. Department of Energy under Contract DE-AC03-76SF00098, and in
part by the National Science Foundation under grant PHY-0098840.

\end{document}